\documentstyle[12pt]{article}

\def\tr{{\rm tr}}
\global\arraycolsep=1pt
\begin{document}

\hfill    SISSA/ISAS 110/94/EP

\hfill    hepth@xxx/9411206

\hfill    August, 1994

\begin{center}
\vspace{10pt}
{\large \bf
NODES AS COMPOSITE OPERATORS \\
IN MATRIX MODELS
\, \footnotemark
\footnotetext{Partially supported by EEC,
Science Project SC1$^*$-CT92-0789.}
}
\vspace{10pt}

{\sl Damiano Anselmi\, \footnotemark\footnotetext{New address after
Sept.\
1$^{st}$, 1994:
Lyman Laboratory, Harvard University, Cambridge MA 02138, U.S.}}

\vspace{4pt}

International School for Advanced Studies (ISAS), via Beirut 2-4,
I-34100 Trieste, Italia\\
and Istituto Nazionale di Fisica Nucleare (INFN) - Sezione di
Trieste,
Trieste, Italia\\

\vspace{12pt}

{\bf Abstract}
\end{center}

\vspace{4pt}

\noindent
Riemann surfaces with nodes
can be described by introducing
simple composite operators in matrix models.
In the case of the Kontsevich model, it is sufficient to add the
quadratic, but ``non-propagating'', term $(\tr[X])^2$ to the
Lagrangian.
The corresponding Jenkins-Strebel differentials
have pairwise identified {\sl simple} poles.
The result is in agreement with a conjecture formulated by Kontsevich
and recently investigated by Arbarello and Cornalba
that the set ${\cal M}_{m*,s}$ of ribbon graphs with $s$ faces and
$m*=(m_0,m_1,\ldots,m_j,\ldots)$ vertices of valencies
$(1,3,\ldots,2j+1,\ldots)$ ``can be expressed in terms of
Mumford-Morita classes'': one gets an interpretation
for {\sl univalent vertices}.
I also address the possible relationship with a recently formulated
theory
of {\sl constrained topological gravity}.
\vfill
\eject

\section{Introduction}
\label{intro}

Strebel's theory of quadratic differentials provides a canonical way
to associate ribbon graphs with metric to Riemann surfaces with
marked points \cite{strebel}.
Such a correspondence can be used to represent the
(decorated) moduli space ${\cal M}_{g,s}$ of Riemann surfaces
of genus $g$ and $s$ marked points as the space of such graphs.
Kontsevich \cite{kontsevich} translated this correspondence into a
matrix
model, that collects the full intersection theory of Mumford-Morita
classes
on ${\cal M}_{g,s}$. The intersection forms are in one-to-one
correspondence
with the physical amplitudes of two dimensional topological gravity
\cite{witten}.

A conjecture
by Kontsevich \cite{kontsevich} states
that the set ${\cal M}_{m*,s}$
of ribbon graphs with $s$ faces and
$m*=(m_0,m_1,\ldots,m_j,\ldots)$
vertices of valencies $(1,3,\ldots,2j+1,\ldots)$
``can be expressed in terms of Mumford-Morita classes''.
This means that there is a deep relation between algebro-geometric
and
combinatoric cohomology classes on the moduli space of Riemann
surfaces \cite{penner}. Recently, the conjecture has been studied in
detail by
Arbarello and Cornalba in \cite{cornalba}. These authors
focus on cases with $m_0=0$.
In this paper, instead, I am concerned with the
geometrical meaning of univalent vertices.

Ribbon graphs with (couples of) univalent vertices
describe Riemann surfaces with nodes.
Precisely, such graphs are constructed with the operators $\tr[X^3]$
and
$(\tr[X])^2$. The former is the same as in the Kontsevich model,
the latter is the ``node''. From the point of view of quadratic
differentials,
the operator $(\tr[X])^2$ stands for a pair of simple poles.

In \cite{preprint,letter} the concept of moduli space constraint in
topological
field theory is
introduced, suggested by supersymmetry \cite{preprint} and naturally
realized
in a field
theoretical model \cite{letter}. ${\cal M}_{g,s}$ is projected onto
the
Pincar\`e dual ${\cal
V}_{g,s}$ of the top Chern class $c_g({\cal E}_{hol})$ of the Hodge
bundle
${\cal
E}_{hol}$, i.e.\ the bundle of abelian differentials. It would be
very
interesting to find the
matrix model counterpart of that constraint. In this paper, we deal
with a
kind of constraint
(nodes) that can be naturally realized in matrix models, but that
seems not so
easily
realizable in a field theoretical framework.

In section \ref{quadr} I recall the main
properties of Jenkins-Strebel quadratic differentials and their role
in the
Kontsevich matrix model. This
facilitates a lot the generalization to Riemann surfaces with nodes,
that
is presented in section \ref{nodes}.
The plumbing fixture can be realized directly ``on the ribbon graph''
and this suggests a deeper relation between
``graphical'' and geometrical concepts.
In section \ref{matrix} I derive the modified matrix model.
Finally, in sect.\ \ref{checks} I exhibit examples of calculations
and provide the interpretation of certain apparently puzzling terms.
Sect.\ \ref{concl} contains the conclusions, where
I address the possible relationship
with a recently formulated theory of ``constrained topological
gravity''
\cite{preprint,letter}.

\section{Quadratic differentials}
\label{quadr}
\setcounter{equation}{0}

It is convenient to recall some properties of quadratic
differentials, before generalizing Strebel's
theorem. In this section, moreover, I briefly recall the Kontsevich
construction of the matrix model describing intersection theory
of Mumford-Morita classes on the moduli space ${\cal M}_{g,s}$
of Riemann surfaces $\Sigma_{g,s}$ of genus $g$ with $s$ marked
points.

Consider a quadratic differential
\begin{equation}
\varphi(z)=\varphi_{zz}(z)dz^2,
\end{equation}
that is a holomorphic
section of the square $(T^*)^{\otimes 2}$ of the tangent
(also called {\sl canonical}) bundle $T^*$.
The trajectories $\gamma=\{z(t),\, t\in{\bf  R}\}$,
along which $\varphi$ is real and positive
\cite{strebel} are called {\sl horizontal trajectories}.
The trajectories where $\varphi$ is real and negative can be called
{\sl
vertical}
trajectories.
Indeed, they are perpendicular to the
horizontal ones:
if $\gamma$ and $\gamma^\prime$
are such that $\varphi(\gamma)>0$ and $\varphi(\gamma^\prime)<0$
and meet at a point $P$, then it is easy to see that ${\rm arg}\left(
\left.{dz\over dt}\right|_P\right)={\rm
arg}\left(\left.{dz^\prime\over dt}
\right|_P\right)\pm{\pi\over 2}$.

A nonzero quadratic differential $\varphi$ defines a
flat metric $ds^2$ on the complement of the (discrete) set of its
zeroes:
\begin{equation}
ds^2=|\varphi(z)||dz|^2.
\label{measur}
\end{equation}
One can measure the length of a horizontal
trajectory: since $\varphi$ is real and positive along
it,
one can take the square root $\sqrt{|\varphi_{zz}|}|dz|$ and
integrate it
\begin{equation}
{\cal L}_{\gamma}=\int_\gamma \sqrt{|\varphi_{zz}|}|dz|=
\int_\gamma ds.
\label{length}
\end{equation}

Let us examine what are the horizontal trajectories
around a pole or a zero. The relevant
behaviours are illustrated in Fig.\ 1.
Let $z=0$ be a zero of order $k$ (or a pole of order $-k$), so that
\begin{equation}
\varphi\sim {\rm e}^{i\alpha}z^kdz^2
\end{equation}
near $z=0$. In polar coordinates $z=\rho{\rm e}^{i\vartheta}$, we
have
\begin{equation}
\varphi\sim \rho^k {\rm e}^{i(k+2)\vartheta+i\alpha}(d\rho+i\rho
d\vartheta)^2.
\label{polar}
\end{equation}
Let us consider rectilinear trajectories entering in the point $z=0$,
namely $\rho=t$, $\vartheta=$const.
Such trajectories will be called {\sl straight}. Then,
$\varphi\sim t^k{\rm e}^{i(k+2)\vartheta+i\alpha}dt^2$ so that
$\varphi>0$
if and only if
\begin{equation}
\vartheta={2n\pi-\alpha\over k+2}.
\end{equation}
Consequently,
there are $k+2$ straight trajectories meeting at $z=0$.
If $k>-2$ the zero is called a {\sl vertex}
and $k+2$ is the {\sl valency} of the vertex. If $k=0$ the two
solutions
$\vartheta=\pi,2\pi$ show that $z=0$ is a {\sl regular point}, while
if $k=-1$
only one trajectory enters the point. For $k<-2$ the situations are
similar to the cases $-k-4$,
as far as straight trajectories are concerned.
Instead, the
case $k=-2$ has to be treated separately. Indeed, for $k=-2$ any
dependence from $\vartheta$ disappears from (\ref{polar}).
If $\alpha=0$ then $\varphi>0$:
any trajectory entering the point is horizontal. If $\alpha\neq 0$
no trajectory entering the point is horizontal.
One can then try with circles winding around $z=0$: $\rho=$const,
$\vartheta=t$. Then $\varphi\sim-\rho^{k+2}{\rm e}^{i\alpha}$, so
that
for $\alpha=\pi$ any circle around the point is horizontal,
independently of its radius $\rho$. In such a case, i.e.\ when
$\varphi\sim -{p^2\over (2\pi)^2}{dz^2\over z^2}$ for some $p\in {\bf
R}_+$,
the area around $z=0$ is
called a {\sl ring domain}.
$p$ is called the {\sl circumference} of the ring domain,
since it is easy to show, using the residue theorem,
that the length (\ref{length})
of the circles of the ring domain
is $p$, independently of the radius $\rho$:
all the circles of a ring domain have the same length.
A ring domain can be thought of as an infinite cylinder,
if one imagines to send the center $z=0$ to infinity.

\setcounter{figure}{0}
\begin{figure}
\begin{picture}(100,500)(0,0)
\end{picture}
\end{figure}

It is easy to find the general solution to the condition
$\varphi>0$, by integrating the differential equation ${\rm Im}
\,\varphi=0$ and by checking the condition
${\rm Re}\, \varphi>0$.
For $k\neq -2$, one finds
\begin{equation}
\left({\rho\over\rho_0}\right)^{k+2}={\sin^2 ({k+2\over 2}
\vartheta_0+{\alpha\over 2})
\over \sin^2 ({k+2\over 2}\vartheta+{\alpha\over 2})}.
\end{equation}
For $k>-2$ the trajectories run away approaching asymptotically the
straight ones: $\rho\rightarrow \infty$
as $\vartheta\rightarrow {2n\pi-\alpha\over k+2}$.
For $k<-2$, on the other hand, the trajectories fall
on the origin, again approaching the straight trajectories, producing
a figure similar to a flower:
$\rho\rightarrow 0$ as $\vartheta\rightarrow {2n\pi-\alpha\over
k+2}$.

For $k=-2$, the solution is
\begin{equation}
\rho=\rho_0{\rm e}^{-{1+\cos\alpha\over \sin \alpha}(\vartheta
-\vartheta_0)}.
\end{equation}
When $\alpha\neq 0,\pi$ one has spirals, $\alpha><0$
determining the orientation.
Illustrative examples are shown in Fig.\ 1 (see also \cite{strebel}).
In the sequel I shall need, in particular, the cases $k=1,-1,-2,-3$.
In the Kontsevich construction one needs $k=1,-2$.

$\varphi$ is called a {\sl Jenkins-Strebel differential} if
the union of nonclosed trajectories has measure zero
(the measure being defined by
$\varphi$ itself according to (\ref{measur})).
In such a case,
the nonclosed trajectories connect the zeroes of $\varphi$ and
draw a graph on the Riemann surface,
decomposing it into ring domains.

A ring domain $R$ is a maximal connected open set
containing no critical point $\varphi$ (i.e.\ no pole and no zero)
and such that \cite{jenkins}:

i) any horizontal trajectory meeting $R$ lies entirely in $R$;

ii) there exists a conformal transformation $w(z)$ mapping $R$ onto
a circular ring $r_1<|w|<r_2$, ($0\leq r_1<r_2$) and mapping the
horizontal trajectories into circles.

When the poles are only simple ones or double poles of the kind
(\ref{behaviour}),
then the ring domains are the only possible maximal domains.
Instead, when there are poles of degree greater than  two
(in which case, however, $\varphi$ is not a Jenkins-Strebel
differential),
there
are also {\sl end domains} (see Fig.\ 1), on which I do not enter in
detail \cite{jenkins}.

The ring domains can be annuli or punctured disks.
Double poles are surrounded by ring domains that are punctured disks.
Annuli can wind around handles.
In the situations I am interested in, there are not annuli but only
punctured disks.

In this section I also assume that the only poles are the
centers of the ring domains.
Let us investigate some properties of the differential $\varphi$ and
its associated graph.
Let $s$ be the number of such
(double) poles, $z_i$ their positions
and $p_i\in{\bf R}_+$, $i=1,\ldots s$ the circumpherences
of the ring domains,
\begin{equation}
\varphi\sim -{p_i^2\over (2\pi)^2}{dz^2\over z^2},\quad \quad
{\rm for}\,\, z\sim z_i.
\label{behaviour}
\end{equation}
The number $n_0$
of zeroes (with multiplicities) of the differential $\varphi$
is easily derived from the degree of $(T^*)^{\otimes 2}$:
\begin{equation}
n_0-2s={\rm deg}[\varphi]=2(2g-2),
\end{equation}
$g$ being the genus of the surface.
Let us assume, for simplicity, that the zeroes are all distinct. Thus
there are $4g-4+2s$ simple zeroes or, equivalently,
the corresponding graph possesses $4g-4+2s$
trivalent vertices.
The faces $F_i$, $i=1,\ldots s$,
of the graph are the maximal ring domains around the
double poles. The $F_i$ are delimited by polygons $P_i$ made of
nonclosed trajectories (the {\sl links} or {\sl edges}
of the graph) connecting the vertices.
``Fatting'' the graph by keeping
some rings close to the polygons $P_i$, one gets a {\sl ribbon graph}
(or {\sl fat graph}). The ribbon graph (with metric,
as we shall see in a moment) identifies the surface unambiguously.

The number $n_1$ of links $\gamma_j$ is easily derived from Euler's
theorem,
\begin{equation}
n_0-n_1+s=2-2g,\quad\quad n_1=6g-6+3s.
\end{equation}
One can associate a length ${\cal L}_j$ to every link $\gamma_j$.
This provides the graph with a metric.
In order for the lengths of the links to be finite, the
Jenkins-Strebel
differential $\varphi$ can have at most double poles and the
double poles must have the form (\ref{behaviour}) in a suitable local
patch.
Since all the rings of a ring domain have the same length,
the perimeter of the polygon delimiting
a face equals the length of its rings.

Now, suppose that the perimeters $p_i$ are fixed and the lengths
${\cal L}_i$ are otherwise arbbitrary.
The number of independent nonzero lengths of the graph
(which will be called the {\sl dimension} of the graph) is
then
\begin{equation}
2(3g-3+s).
\label{1.12}
\end{equation}

If some vertices have valency greater than 3, then the dimension
of the graph is correspondingly reduced.
Expression (\ref{1.12})
equals the (real) dimension of the moduli space ${\cal M}_{g,s}$
of Riemann surfaces $\Sigma_{g,s}$ of genus $g$ with $s$ marked
points.
As a matter of fact, there is a one-to-one correspondence between
Riemann surfaces $\Sigma_{g,s}$ and
{\sl Jenkins-Strebel differentials} of the above kind,
so that the link lengths (under the constraints of fixed
perimeters), together with the graph topologies,
parametrize the moduli space of Riemann surfaces.
Strebel's theorem \cite{strebel,kontsevich}
states that

{\it given a Riemann surface $\Sigma_{g,s}$ with genus $g$ and
$s$ (distinct)
marked points
$z_i$, $i=1,\ldots s$,
and given $s$ positive real numbers $p_i\in {\bf R}_+$, there exists
one
and only one Jenkins-Strebel differential on $\Sigma_{g,s}\backslash
\{z_1,\ldots,z_s\}$, whose maximal ring domains are
punctured disks surrounding the points $z_i$
with circumpherences $p_i$.}

The behaviour of a quadratic differential around a
double pole is an invariant {\sl datum}, i.e.\ it is independent
of the coordinate patch. As a matter of fact, it is the {\sl only}
invariant datum.
Instead, the coefficients of the other poles and the zeroes
can always be set equal to one by a change of local coordinates.
This explains (Fig.\ 1) why the ${1\over z^2}$-coefficient of a
double pole
is so crucial for the qualitative
behaviour of the horizontal trajectories. Instead,
the behaviours around zeroes or other poles are unique.
Strebel's theorem says that
after specifying the surface and the invariant
data, there is one and only one quadratic differential
of the Jenkins-Strebel kind,
that has only double poles and only punctured disks as
maximal ring domains\footnotemark
\footnotetext{Instead, the space of
all quadratic differentials with given poles
$z_i$
of degrees $k_j$ has complex dimension $\sum_jk_j$
(the number of coefficients of the powers
$1/ (z-z_j)^{n_j}$, $n_j=1,\ldots k_j$).}.
So, all the remaining data (positions of the zeroes, coefficients
of the zeroes in a given set of coordinate patches, and so on)
are then fixed. They can be viewed as functions of moduli and
perimeters,
or, equivalently, of the full set of link lengths and graph
topologies.
This remark stresses the deep meaning of the quadratic differentials
we are concerned with. In the next section, I shall show that
when $\varphi$ is allowed to have simple poles, one can only fix
the positions of these simple poles arbitrarily (but not the residue)
and then
the
differential is still unique.

As noticed by Kontsevich \cite{kontsevich}, the correspondence
between Riemann surfaces and ribbon graphs contained in Strebel's
theorem
is one-to-one, namely given an equivalence class
of ribbon graphs, one can easily construct a
surface that admits a corresponding Jenkins-Strebel differential
of the above kind.
This is achieved by filling the faces of the graph with infinite
cylinders,
thus obtaining a surface where the $s$ marked points are sent to
infinity.

\setcounter{figure}{1}
\begin{figure}
\begin{center}
\caption{The propagator ${\rm tr}[\Lambda X^2]$.}
\begin{picture}(200,25)(0,0)
\thicklines
\put(20,12){\vector(1,0){40}}
\put(60,12){\line(1,0){44}}
\put(104,12){\vector(1,0){40}}
\put(144,12){\line(1,0){36}}
\put(20,8){\line(1,0){36}}
\put(96,8){\vector(-1,0){40}}
\put(96,8){\line(1,0){44}}
\put(180,8){\vector(-1,0){40}}
\end{picture}
\end{center}
\end{figure}
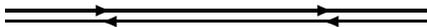

The moduli space of Riemann surfaces can thus be conveniently
described as the space of ribbon graphs.
Such graphs can be
interpreted as Feynmann diagrams of a matrix model
\cite{kontsevich}.
I shall not rederive the Kontsevich matrix model here, since
some more words will be spent
on its generalization to the case of nodes.
Let me simply recall the result and sketch the proof.
Technically, the constraint of fixed perimeters
$p_i$ is overcome by making a Laplace
transform.
This produces an integration over the (flat) space of all link
lengths,
without constraints. Such an integral is
combined with a sum over all kinds of graphs (since
warying the $p_i$ all the graph-topologies are spanned).
The variables $\lambda_i$, dual to $p_i$ in the
sense of Laplace transform, correspond to the entries of a matrix
$\Lambda$ appearing in the quadratic part $\tr [\Lambda
X^2]$ of the matrix model Lagrangian.
The only interaction that is needed is $\tr[X^3]$,
since only trivalent graphs contribute. This is because any graph
with
vertices having valency greater than $3$ can be seen as the limit of
a
trivalent graph when some link lengths tend to zero. This subset is
negligible, in the sense that it has measure zero, the measure being
the product of Mumford-Morita classes, expressed
in terms of link lengths and ring domain perimeters \cite{kontsevich}
(see section \ref{matrix}).
So, the Kontsevich matrix model integral is
\begin{equation}
Z[\Lambda]=c_\Lambda \int dX \,
\exp\left({-{1\over 2}\tr [\Lambda X^2]+{i\over 6}
\tr[X^3]}\right),
\label{kont}
\end{equation}
$c_\Lambda$ being a normalization factor such that the Gaussian
integral obtained neglecting the cubic interaction equals unity.
In Fig.s 2 and 3, the graphical representations of $\tr[\Lambda X^2]$
and $\tr[X^3]$ are shown.

The relation among intersection numbers of Mumford-Morita classes and
Feynmann graphs is more explicitly expressed by the identity
\cite{kontsevich}
\begin{equation}
\sum_{\mbox{\scriptsize{\it
$\sum_{i=1}^sd_i=3g-3+s$}}}<\prod_{i=1}^s \sigma_{d_i}>
\prod_{i=1}^s{(2d_i-1)!!\over \lambda_i^{2d_i+1}}=
\sum_{\Gamma\in G_{g,s}}{2^{-V(\Gamma)}\over S(\Gamma)}\prod_{e\in
L(\Gamma)}{2\over
\tilde\lambda(e)}.
\label{kont2}
\end{equation}
$\sigma_{d_i}=[c_1({\cal L}_i)]^{d_i}$ are the Mumford-Morita
classes,
${\cal L}_i$ being the line bundles on the moduli space
$\bar {\cal M}_{g,s}$, whose fibers are
$T^*_{z_i}\Sigma_{g,s}$. $<\prod_{i=1}^s \sigma_{d_i}>$
stands for
\begin{equation}
\int_{\bar {\cal M}_{g,s}}\prod_{i=1}^s[c_1({\cal L}_i)]^{d_i}.
\end{equation}
$G_{g,s}$ is the set of ribbon graphs $\Gamma$ with $s$ faces and
genus $g$.
$V(\Gamma)$ is the number of vertices of $\Gamma$, while $S(\Gamma)$
is the combinatorial factor, equal to the cardinality of the set of
automorphisms of $\Gamma$. $L(\Gamma)$ denotes the set of links $e$.

The $\lambda$'s are numbers associated to the faces, one for each
face.
Any link $e$ separates two faces; call $e_{ij}$ a link
between the $i^{th}$
and the $j^{th}$ face ($i$ can be equal to $j$).
$\tilde\lambda (e)$ is defined by
$\tilde\lambda(e_{ij})=\lambda_i+\lambda_j$.
${2\over \tilde\lambda(e)}$ is the propagator, while $2^{-V(\Gamma)}$
stands for a factor ${1\over 2}$ for each vertex. The $\lambda$'s
are related to the eigenvalues of the matrix $\Lambda$ appearing in
(\ref{kont}).

\setcounter{figure}{2}
\begin{figure}
\begin{center}
\caption{The vertex operator ${\rm tr}[X^3]$.}
\begin{picture}(196,85)(0,0)
\thicklines
\put(20,42){\vector(1,0){40}}
\put(60,42){\line(1,0){36}}
\put(20,38){\line(1,0){36}}
\put(96,38){\vector(-1,0){40}}
\put(96,42){\vector(2,1){37.8}}
\put(133.8,60.9){\line(2,1){32.2}}
\put(100,40){\line(2,1){32.2}}
\put(168,74){\vector(-2,-1){35.8}}
\put(100,40){\vector(2,-1){35.8}}
\put(135.8,22.1){\line(2,-1){32.2}}
\put(96,38){\line(2,-1){34.2}}
\put(166,3){\vector(-2,1){35.8}}
\end{picture}
\end{center}
\end{figure}
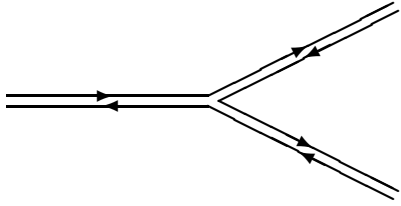

\section{Plumbing fixture for ribbon graphs}
\label{nodes}
\setcounter{equation}{0}

More descriptions of nodes
are analysed in this chapter, the most convenient one leading to
a simple modification of the matrix model (\ref{kont}).
The final result is
\begin{equation}
Z[\Lambda,\lambda]={\rm e}^{W[\Lambda,\lambda]}
=c_\Lambda \int dX \,{\exp}\left({-{1\over 2}\tr [\Lambda X^2]
-{\lambda\over 2}(\tr[X])^2+{i\over 6}\tr[X^3]}\right).
\label{compmodel}
\end{equation}
Amplitudes of order $\lambda^n$ correspond to $n$ nodes.

The function $W[\Lambda,\lambda]$ only contains connected graphs.
{}From Fig.\ 4, which is the graphical representation
of the composite operator $(\tr[X])^2$, we see that one can define
a cutting operation, consisting in the separation of the two
branches $\tr[X]$.
I shall call {\sl reducible} the graphs that become disconnected
when cutting all the $(\tr[X])^2$-insertions
in this way.
The other graphs will be called {\sl irreducible}\footnotemark
\footnotetext{Notice that, instead,
in quantum field theory the {\sl one particle
irreducible} diagrams are those that become disconnected when
cutting a {\sl propagator}.}.
Irreducible connected graphs correspond to pinching non-separating
cycles,
while
reducible connected graphs correspond to pinching a
separating set of cycles.

The only difference with respect to the Kontsevich
matrix model (\ref{kont}) is the introduction of the simplest
composite operator one can imagine, namely $(\tr[X])^2$. I call it
{\sl composite}, since it is the product of operators of the form
$\tr[X^n]$, which I instead call {\sl simple} operators.
{}From the introductory remarks
it follows that the simple operator
$\tr[X^n]$ describes a vertex with $n$ legs of the ribbon graph
or a zero of order $n-2$ and valency $n$ of the
Jenkins-Strebel differential. Thus, no simple operator can describe
nodes. Instead, all composite operators $\tr[X^{n_1}]\,\tr[X^{n_2}]
\cdots$ describe nodes.
The {\sl product} means {\sl identification} of points and
indeed any node is associated with identification of points.
The product $\tr[X^n]\tr[X^m]$ stands for the identification of
a vertex of valency $n$ with a vertex of valency $m$.
$\tr[X]$ corresponds to a valency one ``vertex'', i.e.\ to a simple
pole
of the Jenkins-Strebel differential. $(\tr[X])^2$ is an
identification of two simple poles.

As previously noted,
in the Kontsevich model only the cubic simple
operator $\tr[X^3]$ matters. Similarly,
in our case, the simplest composite operator $(\tr[X])^2$ is
sufficient to describe intersection theory of Mumford-Morita classes
on the moduli space of surfaces with nodes:
the more complicated situations are of zero measure.
Thinking of the graphical representation of $\tr[\Lambda X^2]$ shown
in Fig.\ 2, it is
clear that it is associated to
a regular point on the ribbon graph. This is indeed the
propagator of the matrix model. The nice feature that I want to
stress is that in matrix models
there exists a different quadratic operator, namely $(\tr[X])^2$,
that can be thought as playing a different role,
instead of {\sl propagating}.
Thinking to its graphical representation
(Fig.\ 4), it is clear that $(\tr[X])^2$
does not represent a regular point, rather it
very plausibly represents a node.

\setcounter{figure}{3}
\begin{figure}
\begin{center}
\caption{The composite operator $({\rm tr}[X])^2$.}
\begin{picture}(200,25)(0,0)
\thicklines
\put(20,12){\vector(1,0){40}}
\put(60,12){\line(1,0){36}}
\put(104,12){\vector(1,0){40}}
\put(144,12){\line(1,0){36}}
\put(96,8){\line(0,1){4}}
\put(104,8){\line(0,1){4}}
\put(20,8){\line(1,0){36}}
\put(96,8){\vector(-1,0){40}}
\put(104,8){\line(1,0){36}}
\put(180,8){\vector(-1,0){40}}
\end{picture}
\end{center}
\end{figure}
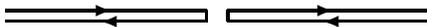

For some time I shall focus on non-separating nodes.
The extension to separating
nodes will be straightforward.

Thus Strebel's theorem should be generalized in the following way.

{\it Given $s$ positive real numbers $p_i\in {\bf R}_+$,
for quasi every a Riemann surface
$\Sigma_{g,s,n}$ with genus $g$, $s$ (distinct) marked points
$z_i$, $i=1,\ldots s,$ and $n$ nodes,
represented by $2n$ pairwise identified marked points
$(x_j,y_j)$, $j=1,\ldots n$,
there exists one
and only one Jenkins-Strebel differential $\varphi$ on
$\Sigma_{g,s,n}
\backslash
\{z_1,\ldots,z_s,x_1,\ldots,x_n,\-
y_1,\ldots,y_n\}$ such that

i) the maximal ring domains are punctured disks, surrounding the
points $z_i$ with circumpherences $p_i$;

ii) $\varphi$ has simple poles in the points $x_j$, and $y_j$.}

The reason for the ``quasi'' is that the theorem holds except for a
set
of vanishing measure in the moduli space. Indeed, letting some link
length
tend to zero, it is always possible to make, say, a zero of order $n$
and a simple pole collapse, thus producing a limiting
situation with a zero of order $n-1$ and one missing simple pole.
Stated differently, after having fixed the positions of the poles
and the numbers $p_i$, the differential $\varphi$ is uniquely fixed.
It has a certain number of zeroes placed in certain positions.
Such positions cannot coincide with the positions of the double
poles,
but can ``coincide with the positions of the simple ones''.

Let us check that the above space of ribbon graphs has the
correct dimension. I begin with the case in which the $n$ nodes are
all
non-separating.
The first thing to do is to find the number $n_1$
of links. By the degree of $\varphi$, the number $n_0$ of zeroes is
now
\begin{equation}
n_0-2s-2n={\rm deg}[\varphi]=2(2g-2), \quad\quad
n_0=4g-4+2s+2n.
\label{n0}
\end{equation}
Supposing that all the zeroes are distinct, the number of links
results
\begin{equation}
n_1={3n_0+2n\over 2}=6g-6+3s+4n.
\label{n1}
\end{equation}
We conclude that, keeping the $s$ perimeters $p_i$ fixed, the
dimension
of the graph is
\begin{equation}
2(3g-3+s+2n)={\rm dim}_{\bf R}{\cal M}_{g,s,n},
\label{3.4}
\end{equation}
${\cal M}_{g,s,n}$ denoting the moduli space of Riemann
surfaces $\Sigma_{g,s,n}$ with $n$ nonseparating nodes.
Indeed, such moduli space is a subspace
of the moduli space ${\cal M}_{g+n,s}$, where $n$ cycles
have been pinched. Now,
${\rm dim}_{\bf R}{\cal M}_{g+n,s}=6g-6+2s+6n$ and each pinching
reduces the real dimension by $2$.

In the case when there are $n$ non-separating nodes and $m$
separating nodes, a similar computation gives $6g-6+2s+4n-2m$, both
for the dimension of the moduli space, which I denote by
${\cal M}_{g,s,n,m}$, and the number of independent link
lengths of the ribbon graph.

The proof of the above generalization of Strebel's theorem is very
simple.

Given a Riemann surface of genus $g$ and $s+2n$ points and given $
s+2n$ real and positive numbers $p_1,\ldots p_s$, $q_1,\ldots q_n$,
$r_1,\ldots r_{n}$, Strebel's theorem associates
a unique quadratic differential to it.
Let us call {\sl virtual} the faces with perimeters
$q_j$ and $r_j$ and the corresponding ring domains.
Let us now take the limit
$q_j,r_j\rightarrow 0$: the virtual faces are shrunk to
points, absorbing some zeroes. The most simple situation
is the one in which any virtual face is surrounded by a single link:
a single zero is made to collapse with the double pole that
centers the virtual face, thus producing a simple pole,
\begin{equation}
\varphi\sim (\alpha z-r^2){dz^2\over z^2}\rightarrow \alpha{dz^2
\over z}.
\end{equation}
The dimension of the graph is unchanged.
One can easily
check that it is the correct value (\ref{3.4}).
In all the more complicated situations,
i.e.\ when at least one virtual face is delimited by two or more
links,
the dimensions of the resulting graphs are smaller than the expected
value.
Consequently, they are a subset of vanishing measure.

The reason why we can take the limit $q_j,r_j\rightarrow 0$ is clear:
the correlation functions are independent of $q_j,r_j$ and moreover,
no Mumford-Morita class is placed in the nodes. Indeed,
the perimeters $p_i$ of the ``true'' marked points
are to be kept different from
zero because they describe the line bundles ${\cal L}_i$:
$p_i\neq 0$ are required to express $c_1({\cal L}_i)$
(see formula (\ref{morita})) and to perform calculations. However,
no obstruction of this kind prevents from taking the above limit
in the case of nodes.
With $q_j,r_j$ different from zero, we have a different
but equivalent description of nodes, that however,
cannot be directly inserted into a matrix model.

I now describe the theorem just proven in terms of some
{\sl plumbing fixture for ribbon graphs}. This discussion could
stimulate
the investigation of the ``graphical counterparts'' of
certain ``geometrical concepts''.

The idea
is that of starting again
from Strebel's thorem and pinching the cycles directly
on the ribbon graph, showing ``dynamically''
that one gets precisely a graph
constructed with the composite operator $(\tr[X])^2$.

Let us consider a surface $\Sigma_{g+n,s}\in{\cal M}_{g+n,s}$.
We know that there exists a unique
Jenkins-Strebel differential $\varphi$ on it,
given the
perimeters $p_1,\ldots p_s$.
Let $\Gamma$ denote the ribbon graph associated with $\varphi$.
Consider the handle that has to be pinched.
The most simple behaviour of $\Gamma$ around the handle
is shown in Fig.\ 5. It will be sufficient,
as we shall see, to restrict to this situation.
So, let us focus on this behaviour, for now.

\setcounter{figure}{4}
\begin{figure}
\begin{center}
\caption{Typical behaviour of a graph around a handle
and scheme of it.}
\begin{picture}(370,80)(0,0)
\thicklines
\put(50,40){\oval(40,60)}
\put(190,40){\oval(40,60)[r]}
\put(142,50){\line(1,0){18}}
\put(210,50){\vector(-1,0){50}}
\put(142,46){\vector(1,0){18}}
\put(160,46){\line(1,0){50}}
\put(142,46){\line(0,-1){16}}
\put(118,50){\oval(40,40)[rt]}
\put(118,30){\oval(40,40)[rb]}
\put(122,50){\oval(40,40)[rt]}
\put(122,30){\oval(40,40)[rb]}
{\thinlines
\put(118,40){\oval(40,60)[l]}
\put(122,40){\oval(40,60)[l]}
}
\put(70,30){\vector(1,0){50}}
\put(120,30){\line(1,0){18}}
\put(138,34){\vector(-1,0){18}}
\put(70,34){\line(1,0){50}}
\put(138,34){\line(0,1){16}}
\put(50,70){\line(1,0){140}}
\put(50,10){\line(1,0){140}}
\put(300,42){\oval(40,40)[t]}
\put(300,42){\oval(32,32)[t]}
\put(300,38){\oval(40,40)[b]}
\put(300,38){\oval(32,32)[b]}
\put(284,42){\vector(0,-1){4}}
\put(320,42){\vector(0,-1){4}}
\put(304,40){\line(6,1){12}}
\put(304,40){\line(6,-1){12}}
\put(268,40){\line(6,1){12}}
\put(268,40){\line(6,-1){12}}
\put(300,40){\makebox(0,0){$0$}}
\put(260,40){\makebox(0,0){$\infty$}}
\end{picture}
\end{center}
\end{figure}
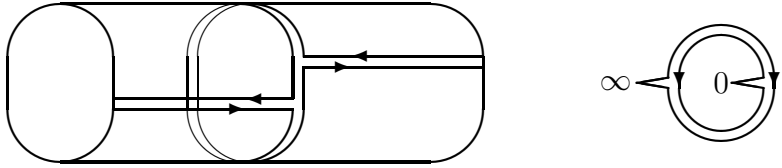

We see that the Jenkins-Strebel differential $\varphi$ possesses
two simple zeroes and no pole around the cycle. Mapping the handle
into a plane we get the structure represented
on the right hand side of Fig.\ 5.
The V's denote, as I show a in moment, poles of order 3 (see also
Fig.\ 1).
Let $q\in {\bf C}$
be such that $|q|$ parametrizes
the size of the handle. $q$ will be made to vanish
when the cycle will be pinched. However, I want to perform a plumbing
fixture such that the two zeroes of the differential $\varphi$
are made to collapse on the node at the same time as the node is
formed.
In other words, the representative $\gamma$ of the
cycle that I want to pinch
is chosen to be a union of nonclosed horizontal trajectories of the
Jenkins-Strebel differential $\varphi$, precisely
those trajectories that connect the above two vertices of the ribbon
graph.
That means that, in a suitable local coordinate frame the
two zeroes have coordinates tending to zero with $q$.
The simplest behaviour of $\varphi$
around $\gamma$ is
\begin{equation}
\varphi\sim \left(\alpha z+ \beta{q^2\over z}\right){dz^2\over z^2},
\label{diff}
\end{equation}
for, say, $|q|^2<|z|<1$, $\alpha,\beta$ being independent of $q$.
The plumbing fixture is realized by identifying $z$ with ${q^2\over
w}$,
$|q|^2<|w|<1$. In the $w$-coordinate system, $\varphi$ looks like
\begin{equation}
\varphi\sim \left(\beta w+\alpha{q^2\over w}\right){dw^2\over w^2}.
\end{equation}

Let us first study the differential on the full plane $0<|z|<\infty$.
There are two simple zeroes, as desired, for $z=\pm
iq\sqrt{\beta/\alpha}$.
Moreover, there are two triple poles, one in $z=0$ and one
at infinity. The general information on quadratic
differentials given in the previous section,
is sufficient to prove that the behaviour
is precisely the one depicted in the scheme of Fig.\ 5.
The triple poles are outside
the region $|q|^2<|z|<1$, while the zeroes and $\gamma$
are inside.

We see that when $\gamma$ is pinched, i.e.\ $|q|\rightarrow 0$, then
$\varphi(z)$ approaches $\alpha{dz^2\over z}$, which has a simple
pole
on the node, as claimed. Similarly, on the other branch of the node
$\varphi(w)\sim \beta{dw^2\over w}$. We also see that, given the
coordinate patches, the residues of the simple poles are
automatically fixed.

The form (\ref{diff}) of $\varphi$ is simplified, since the two
zeroes
$z=\pm iq$ have symmetric positions. In general one has a behaviour
like
\begin{equation}
\varphi\sim \left(\alpha z+\beta q+\gamma {q^2\over
z}\right){dz^2\over z^2}
=\left(\gamma w+\beta q +\alpha {q^2\over w}\right){dw^2\over w^2},
\end{equation}
and the previous conclusions still hold.
$\alpha$, $\beta$ and $\gamma$ are functions of the moduli,
(since $\varphi$ is unique), with a finite limit as $q\rightarrow 0$.
The most important fact is that the lengths of the two edges
forming the cycle $\gamma$ tend to zero contemporarily and
with the same velocity. Precisely, they tend to zero as $|q|^{1/2}$
(this fact follows from a simple dimensional analysis).

Note that the plumbing fixture was realized with $xy=q^2$ and
not, for example, $xy=t$. With $xy=t$, one would be forced to
introduce the square root of $t$.

Let us now discuss the situations that are more complicated than
Fig.\ 5.
One can choose to pinch any closed path $\gamma$ on the ribbon graph
which is made of the union of nonclosed trajectories and that lies in
the
homology class of the given
non-separating cycle. Let $k$ be the number of
zeroes (with multiplicities) lying on $\gamma$. It is clear that
$k\geq 2$. Indeed, one simple zero ($k=1$) is not possible.
Such zero
corresponds to a trivalent vertex. Two of the three legs of this
vertex
lie on $\gamma$, while the remaining one lies is one (say the left
one)
of the two branches in which
the handle is sectioned when cutting along $\gamma$. This means that
there is
no leg
exiting
from $\gamma$ and lying on the right branch of the handle.

So, such side is a ring domain. However, since it cannot be an
annulus,
due to Strebel's theorem
(the ring domains of $\varphi$ are
only punctured disks), it can only contain a single puncture.
This means that the right hand side of the handle is a {\sl face}
in the ribbon graph, and has the path $\gamma$ as a boundary.
As a matter of fact, such (separating) cycles cannot be pinched. The
``graphical'' reason is that the length of $\gamma$ is a perimeter
and thus
is fixed. The corresponding ``geometrical''
reason is that such a pinching would produce a sphere with
two marked points, one being the puncture and the other being the
node.
Such a surface has nontrivial isometries, since it is necessary
to mark at least three
points on the sphere to eliminate isometries, due to $SL(2,{\bf C})$
invariance. These situations are, in some sense, ``set to zero''.
If $k=2$, $\gamma$
is the union of two nonclosed trajectories,
at least when the zeroes situated on
$\gamma$ are simple. Their lengths are set to
zero when pinching. This reduces the number of moduli by $2$, as
desired. When $k>2$, on the other hand, so that $\gamma$ is
the union of $k$ nonclosed trajectories, one looses $k$ lengths when
pinching. The final ribbon graph is constructed with composite
operators of the form $\tr[X^m]\tr[X^n]$, but it
can always be seen as the particular
case of a suitable ribbon graph constructed with $\tr[X^3]$ and
$(\tr[X
])^2$ where the lengths of some links are set to zero.
To be explicit, an example of the behaviour of $\varphi$
for generic $k$ is
\begin{equation}
\varphi\sim \left(\sum_{n=-k}^k\alpha_n z^nq^{k-n}\right){dz^2\over
z^2}
\sim \left(\sum_{n=-k}^k\alpha_{-n}w^nq^{k-n}\right){dw^2\over w^2}.
\end{equation}
The subset of ribbon graphs corresponding to these pinchings
is negligible, i.e.\ it is of zero
measure, since the wedge of
Mumford-Morita classes is regular when some lengths are set to zero.
Again, this explains the ``quasi'' of the theorem and justifies the
generality of the set of ribbon graphs constructed {\sl via} the
composite
operator $(\tr[X])^2$.

Let us now describe another kind of pinching, that is directly
related to the proof of the theorem of this section. The idea is to
shrink $\gamma$ together with {\sl one} of its zeroes, not both.
This can be obtained by introducing an arbitrary $r\in {\bf R}_+$ and
defining
$q$ so that
\begin{equation}
\varphi\sim \left(\alpha z-r^2+
\beta q+\gamma {q^2\over z}\right){dz^2\over z^2}
=\left(\gamma w-r^2+\beta q +\alpha {q^2\over w}\right){dw^2\over
w^2}.
\end{equation}
In this way, only one of the two
zeroes falls on the triple pole, as desired.
Consequently, a double pole is produced
instead of a simple one. Notice that the perimeters of the two
identified double poles are equal (instead, a similar statement
is not meaningful for
the residues of the identified simple poles, since such residues
depend on the coordinate patches).

The final task is to find the precise
correspondence between intersection forms on ${\cal M}_{g,s,n}$ and
the matrix model (\ref{compmodel}).

\section{Matrix model}
\label{matrix}
\setcounter{equation}{0}

The Mumford-Morita classes are easily expressed in terms of link
lengths and perimeters. Let us consider a certain ring domain, of
perimeter $p$ and let $l_1,\ldots l_k$ be the lengths of the links
delimiting it, $\sum_{i=1}^kl_i=p$.
Then the corresponding Mumford-Morita class $\omega$
reads \cite{kontsevich}
\begin{equation}
\omega=\sum_{\mbox{\scriptsize{\it $1\leq i\leq j\leq k-1$}}}
d\left({l_i\over p}\right)\wedge d\left({l_j\over p}\right).
\label{morita}
\end{equation}
It is clear that such an expression also holds in our case, where the
``polygon'' delimiting the face can be of the form shown in Fig.\
6. Indeed, (\ref{morita}) surely holds before pinching. Since
our pinching corresponds to set some link lengths to zero,
(\ref{morita}) also holds after pinching.

The correlation functions can be expressed as
\begin{equation}
\int_{\bar {\cal M}_{g,s,n}}\prod_{i=1}^s[c_1({\cal L}_i)]^{d_i}=
<\prod_{i=1}^s\sigma_{d_i}>_n=\int_{\pi_n^{-1}(p*)}
\prod_{i=1}^n\omega_i^{d_i}.
\label{4.2}
\end{equation}
$p*=(p_1,\ldots,p_s)$ is the sequence of perimeters, $\pi_n^{-1}(p*)$
denotes the space of irreducible connected
ribbon graphs with fixed perimeters and $n$
insertions of the composite operator $(\tr[X])^2$.
The intersection numbers are independent of the values $p_i$.

The proof that the amplitudes (\ref{4.2}) are encoded in the matrix
model
(\ref{compmodel}) goes on along the same lines of the proof made in
\cite{kontsevich}. Nevertheless, it is worth making the main steps
explicitly.

Let us define the form \cite{kontsevich}
\begin{equation}
\Omega=\sum_{i=1}^sp_i^2\omega_i.
\end{equation}
$\Omega$ is useful, since I can collect all the amplitudes
with the same value of $d=\sum_{i=1}^sd_i$ in the
expression
\begin{equation}
V_n(p*)=\int_{\pi_n^{-1}(p*)}{\Omega^d\over d!}=
\sum_{\mbox{\scriptsize{\it $\sum_{i=1}^sd_i=d$}}}
 <\prod_{i=1}^s\sigma_{d_i}>_n\prod_{i=1}^s{p_i^{2d_i}\over d_i!}.
\end{equation}
In order for this expression to be nonzero, $d$ must be equal
to ${\rm dim}_{\bf C}{\cal M}_{g,s,n}=3g-3+s+2n$.

The integration over $\pi_n^{-1}(p*)$ is facilitated if we also
integrate over $p*$. In order to keep all information, we make a
Laplace transform in the perimeters $p_i$, namely
\begin{equation}
\int \prod_{i=1}^sdp_i \,{\rm e}^{-\sum_{i=1}^s\lambda_ip_i}.
\end{equation}
After the Laplace transform, one finds, on the right hand side,
an integration over the space
of ribbon graphs. Such integration
can be split into a sum $\sum_{\Gamma\in G_{g,s,n}}$
over the types of graphs, wheighted by
the usual combinatorial factor ${1\over S(\Gamma)}$,
times the integration over the lengths of the edges of the graph.
Let us denote by $\rho_{g,s,n}$
the Jacobian factor that arises when changing
the integration variables in this way. One gets
\begin{equation}
\sum_{\mbox{\scriptsize{\it $\sum_{i=1}^sd_i=d$}}}
\, <\prod_{i=1}^s\sigma_{d_i}>_n\prod_{i=1}^s
{(2d_i)!\over d_i!\lambda_i^{2d_i+1}}=
\sum_{\Gamma\in{G}_{g,s,n}}
{1\over S(\Gamma)}\int_{{\bf R}_+^{L(\Gamma)}}
\rho_{g,n,s}\,{\rm e}^{-\sum_{i=1}^s\lambda_ip_i}
\prod_{e\in L(\Gamma)}dl(e),
\label{hform}
\end{equation}
$L(\Gamma)$ denoting both
the set of links $e$ of the graph $\Gamma$ and the cardinality of
this set, given in (\ref{n1}) (no confusion can arise).
One has, by definition,
\begin{equation}
\rho_{g,n,s}=\left|{\left(\prod_{i=1}^sdp_i\right){\Omega^d\over d!}
\over \prod_{e\in L(\Gamma)}dl(e)}\right|.
\label{rho}
\end{equation}
Clearly, as in \cite{kontsevich}, $\rho_{g,n,s}$ is
locally constant and only depends on the combinatorial type of the
graph. Indeed, at fixed perimeters, $\Omega$ is simply a sum
of products of differentials of link lengths. We expect that an
expression
very similar to the one of ref.\ \cite{kontsevich} holds, namely
\begin{equation}
\rho_{g,s,n}=2^{d+L(\Gamma)-V(\Gamma)}=2^{5g-5+2s+4n},
\label{4.8}
\end{equation}
where $V(\Gamma)$ is the number of trivalent vertices of the graph,
given in (\ref{n0}),
while $L(\Gamma)$ is the number of links, given by (\ref{n1}).
As a matter of fact, on pag.\ 12 of ref.\ \cite{kontsevich}
one can find the general formula
\begin{equation}
\rho=4^D 2^{1-g},\quad \quad D={1\over 2}({\rm dim}\, {\cal
M}_{m*,s}-
s),
\end{equation}
${\rm dim}\, {\cal M}_{m*,n}$ being the number of edges of the graphs
made with $m*=(m_0,m_1,\ldots,m_j,\ldots)$ vertices of valencies
$(1,3,\ldots,2j+1,\ldots)$.
Setting $m*=(2n,n_0,0,\ldots)$, one arrives at (\ref{4.8}),
$D$ being a half of (\ref{3.4}).
I have checked formula (\ref{4.8})
explicitly, by direct application
of the definition (\ref{rho}), in all the explicit
computations that will be presented in the next
section.

The integration over $l(e)$ in the right hand side of (\ref{hform})
is the same as in \cite{kontsevich}.
At the end one finds
\begin{equation}
\sum_{\mbox{\scriptsize{\it
$\sum_{i=1}^sd_i=3g-3+s+2n$}}}<\prod_{i=1}^s \sigma_{d_i}>_n
\prod_{i=1}^s{(2d_i-1)!!\over \lambda_i^{2d_i+1}}=
\sum_{\Gamma\in G_{g,s,n}}{2^{-V(\Gamma)}\over S(\Gamma)}\prod_{e\in
L(\Gamma)}{2\over
\tilde\lambda(e)}.
\label{mai}
\end{equation}

Let $\Lambda={\rm diag}(\Lambda_1,\ldots,\Lambda_N)$ be an $N\times
N$
positive hermitian diagonal matrix ($N$ will be let tend to infinity)
and let
\begin{equation}
t_i(\Lambda)=-(2i-1)!!\,\,\tr[\Lambda^{-(2i+1)}].
\end{equation}
I define the generating function of the amplitudes as
\begin{equation}
F(\lambda,t_0,t_1,\ldots)=
\sum_n{\lambda^n}\sum_{(k)}
<\prod_{i=0}^\infty\sigma_i^{k_i}>_n \prod_{j=0}^\infty
{t_j^{k_j}\over k_j!}.
\end{equation}
Now, we can write
\begin{eqnarray}
F(\lambda,t_0(\Lambda),t_1(\Lambda),\ldots)&=&\sum_n{\lambda^n}
\sum_{s>0}\sum_{d_i\geq 0}<\prod_{i=1}^s\sigma_{d_i}>_n
{1\over s!}\prod_{j=1}^st_{d_j}(\Lambda)\nonumber\\
&=&\sum_n{\lambda^n}
\sum_{s>0}{(-1)^s\over s!}\sum_{d_i\geq 0}
<\prod_{i=1}^s\sigma_{d_i}>_n
\prod_{i=1}^s(2d_i-1)!!\sum_{j=1}^N{1\over
\Lambda_j^{2d_i+1}}\nonumber\\
&=&\sum_n{\lambda^n}
\sum_{s>0}{(-1)^s\over s!}\sum_{\mbox{
\scriptsize{\it $1\leq j_1,\cdots j_s\leq N$}}}
\sum_{d_i\geq 0}
<\prod_{i=1}^s\sigma_{d_i}>_n
\prod_{i=1}^s{(2d_i-1)!!\over \Lambda_{j_i}^{2d_i+1}}.
\nonumber\\
\end{eqnarray}
Using (\ref{mai}) we arrive at
\begin{equation}
F(\lambda,t_0(\Lambda),t_1(\Lambda),\ldots)=
\sum_n{(-\lambda)^n}\sum_{s>0}\sum_{\Gamma\in
G_{n,s,N}}\left({i\over 2}\right)^{V(\Gamma)}{1\over S(\Gamma)}
\prod_{e\in L(\Gamma)}{2\over \tilde\Lambda(e)}.
\label{fin}
\end{equation}
Here $G_{n,s,N}$ denotes the set of irreducible graphs with $n$
non-separating pinched cycles, $s$ faces and $N$ eigenvalues
$\Lambda_j$
assigned to the faces. The right hand side of (\ref{fin}) is also
the collection of the irreducible connected
Feynamm graphs of the matrix model
(\ref{compmodel}).
We conclude that $F(\lambda,t_0(\Lambda),t_1(\Lambda),\ldots)$
is an asymptotic expansion for the matrix model (\ref{compmodel})
as $\Lambda^{-1},\lambda\rightarrow 0$.

In the derivation I have always considered non-separating nodes, but
the final expression (\ref{compmodel}) clearly collects also diagrams
corresponding to the pinching of separating cycles
(reducible graphs).

\section{Checks, calculations and comments}
\label{checks}
\setcounter{equation}{0}

Using a theorem of \cite{difra}, I prove that the amplitudes of
(\ref{compmodel})
agree with their geometrical interpretation. This will also permit to
clarify
some subtelties. Let us begin with some explicit examples.
The comparison between left and right hand sides of formula
(\ref{mai})
gives
\begin{equation}
\matrix{<\sigma_0>_1={1\over 2},&
<\sigma_2>_2={1\over 8},&
<\sigma_4>_3={1\over 48},\cr
<\sigma_0\sigma_1>_1={1\over 2},&
<\sigma_0\sigma_0\sigma_2>_1={1\over 2},&
<\sigma_0\sigma_1\sigma_1>_1=1,\cr
<\sigma_0\sigma_3>_2={1\over 8},&
<\sigma_1\sigma_2>_2={3\over 8},&\cr
<\sigma_0\sigma_5>_3={1\over 48},&
<\sigma_1\sigma_4>_3={5\over 48},&
<\sigma_2\sigma_3>_3={5\over 24}.}
\label{example}
\end{equation}
The calculation of
$<\sigma_0\sigma_1>_1$ involves the sum of three graphs
(not counting the permutations of faces),
$<\sigma_0\sigma_0\sigma_2>_1$ and
$<\sigma_0\sigma_1\sigma_1>_1$ require the sum of 24 graphs,
while
$<\sigma_0\sigma_3>_2$ and $<\sigma_1\sigma_2>_2$
involve the sum of 21 graphs. Finally,
$<\sigma_0\sigma_5>_3$,
$<\sigma_1\sigma_4>_3$ and
$<\sigma_2\sigma_3>_3$ are found by summing 48 graphs.

To be explicit, in Fig.\ 6 I collect the three graphs of
$<\sigma_0\sigma_1>_1$. We have
\begin{equation}
\Gamma_1={1\over \lambda_1}{1\over \lambda_2}\left({2\over\lambda_1+
\lambda_2}\right)^2,\quad
\Gamma_2={1\over 2}{1\over \lambda_1^2}\left({2\over \lambda_1+
\lambda_2}\right)^2,\quad
\Gamma_3={1\over \lambda_1^3}{2\over \lambda_1+\lambda_2}.
\end{equation}
The only graph having nontrivial automorphisms is clearly $\Gamma_2$.
This is the reason for the factor ${1\over 2}={1\over S(\Gamma_2)}$.
Formula (\ref{mai}) gives then
\begin{equation}
<\sigma_0\sigma_1>_1\left({1\over\lambda_1\lambda_2^3}
+{1\over \lambda_1^3\lambda_2}\right)={1\over 4}\Gamma_1+
{1\over 4}(
\Gamma_2+\Gamma_3+\lambda_1\leftrightarrow\lambda_2)=
{1\over 2}{\lambda_1^2+\lambda_2^2\over \lambda_1^3\lambda_2^3}.
\end{equation}

\setcounter{figure}{5}
\begin{figure}
\begin{center}
\caption{Graphs contributing to $<\sigma_0\sigma_1>_1$.}
\begin{picture}(276,75)(10,-10)
\thicklines
\put(70,32){\oval(40,40)[t]}
\put(70,32){\oval(32,32)[t]}
\put(70,28){\oval(40,40)[b]}
\put(70,28){\oval(32,32)[b]}
\put(54,32){\vector(0,-1){4}}
\put(90,32){\vector(0,-1){4}}
\put(74,32){\line(1,0){12}}
\put(74,28){\line(1,0){12}}
\put(74,28){\line(0,1){4}}
\put(38,32){\line(1,0){12}}
\put(38,28){\line(1,0){12}}
\put(38,28){\line(0,1){4}}
\put(142,32){\oval(40,40)[t]}
\put(142,32){\oval(32,32)[t]}
\put(142,28){\oval(40,40)[b]}
\put(142,28){\oval(32,32)[b]}
\put(122,32){\vector(0,-1){4}}
\put(162,28){\vector(0,1){4}}
\put(146,32){\line(1,0){12}}
\put(146,28){\line(1,0){12}}
\put(146,28){\line(0,1){4}}
\put(126,32){\line(1,0){12}}
\put(126,28){\line(1,0){12}}
\put(138,28){\line(0,1){4}}
\put(214,32){\oval(40,40)[t]}
\put(214,32){\oval(32,32)[t]}
\put(214,28){\oval(40,40)[b]}
\put(214,28){\oval(32,32)[b]}
\put(194,32){\vector(0,-1){4}}
\put(198,28){\vector(0,1){4}}
\put(234,32){\line(1,0){18}}
\put(234,28){\line(1,0){18}}
\put(230,32){\line(0,-1){4}}
\put(252,32){\line(0,1){12}}
\put(252,28){\line(0,-1){12}}
\put(256,32){\line(0,1){12}}
\put(256,28){\line(0,-1){12}}
\put(256,32){\line(0,-1){4}}
\put(252,44){\line(1,0){4}}
\put(252,16){\line(1,0){4}}
\put(198,32){\line(0,-1){4}}
\put(70,-5){\makebox(0,0){$\Gamma_1$}}
\put(142,-5){\makebox(0,0){$\Gamma_2$}}
\put(214,-5){\makebox(0,0){$\Gamma_3$}}
\end{picture}
\end{center}
\end{figure}
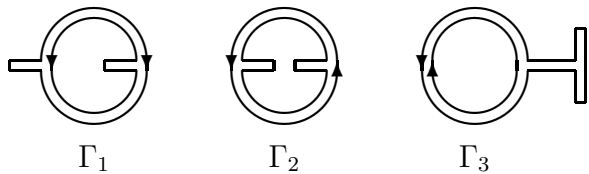

All the computations (\ref{example})
are in genus zero. An ``empirical rule''
for explaining the indicated values is the following.
In all the cases (\ref{example}) we
have
\begin{equation}
<\prod_{i=1}^s\sigma_{d_i}>^{irr}_n={1\over 2^n n!}{(s+2n-3)!\over
\prod_{i=1}^s
d_i!},
\label{for}
\end{equation}
with $\sum_{i=1}^sd_i=-3+s+2n$. $<\ldots>_n^{irr}$
denotes the set of irreducible graphs.
This formula can be understood as follows.
Each pinched
cycle is represented by a couple of identified marked points.
The operator $\sigma_0$ has indeed the simple effect of marking one
point,
so that each node can be described as a couple $\sigma_0\sigma_0$.
Moreover a factor one half for each node is due to the
identifications of the points of each couple. Finally,
when more nodes are present, the overall
factor ${1\over n!}$ takes care of the identity of each couple of
points.

Due to this, we expect
\begin{equation}
<\prod_{i=1}^s\sigma_{d_i}>^{irr}_n
=<{1\over n!}\left({\sigma_0\sigma_0\over 2}\right)^n
\prod_{i=1}^s\sigma_{d_i}>_0,
\label{argue}
\end{equation}
which reduces to (\ref{for}) for $g=0$.

Formula (\ref{argue}) can be proved using the results of ref.\
\cite{difra}. This is a check of the treatment of nodes
within matrix models. A clarifying subtlety will come out.
In ref.\ \cite{difra} it is proved that for every polynomial
$P(X)$ in the traces of the odd powers of $X$, there exists a
polynomial $Q\left({\partial\over \partial t}\right)$ in the
derivatives with respect to the $t_i$'s such that
\begin{equation}
\ll P(X)\gg
=Q\left({\partial\over \partial t}\right)
\ll 1 \gg,
\label{yto}
\end{equation}
and {\sl vice versa},
where
\begin{equation}
\ll P(X)\gg\equiv
c_\Lambda\int dX\,P(X)\,
{\rm exp}\left(-{1\over 2}\tr[\Lambda X^2]+{i\over 6}\tr[X^3]\right).
\end{equation}
In particular, the following result holds:
\begin{equation}
{1\over k!}{\partial^k\over\partial t_0^k}\ll 1\gg=
\sum_{\mbox{\scriptsize{\it $m,n\geq 0,\,\, 3n+m=k$}}}
{(-i)^m\over m!\, n!\, 6^n}
\ll (\tr[X])^m\gg.
\label{equa}
\end{equation}
 From this formula we derive
\begin{equation}
\ll (\tr[X])^{2n}\gg=(-1)^n{\partial^{2n}\over \partial t_0^{2n}}
\ll 1\gg+\cdots,
\label{popu}
\end{equation}
the dots standing for derivatives of the kind $\left({\partial
\over \partial t_0}\right)^{2n-3m}$, $m>0$. The subtlety addressed to
at the beginning of this section
has to do with the interpretation of these apparently puzzling extra
terms.

One has
\begin{equation}
<\prod_{i=0}^\infty\sigma_i^{k_i}>_n=
{1\over n!}\left(-{1\over 2}\right)^n
\left.\prod_{i=0}^\infty{\partial^{k_i}\over \partial t_i^{k_i}}
\ll (\tr[X])^{2n}\gg_c\right|_{t=0},
\end{equation}
where $\ll \ldots \gg_c$ denotes the set of connected
graphs\footnotemark
\footnotetext{In defining connectedness, $(\tr[X])^{2n}$
has to be considered as a set of $n$ distinct operators
$(\tr[X])^{2}$.}, namely
\begin{equation}
\ll (\tr[X])^{2n}\gg_c\equiv
{\partial^n\over \partial h^n}\left.
\ln\left(\sum_{k=0}^\infty
{h^k\over k!}\ll (\tr[X])^{2k}\gg\right)\right|_{h=0}.
\end{equation}
{}From (\ref{popu}), one gets
\begin{equation}
\ll (\tr[X])^{2n}\gg_c=(-1)^n{\partial^{2n}\over \partial t_0^{2n}}
\ll 1\gg_c+\cdots,
\label{popu2}
\end{equation}
where now the dots also contain terms like
$\prod_j\left({\partial^{n_j}\over \partial t_0^{n_j}}\ll 1
\gg_c\right)$
with $\sum_j n_j=2n,2n-3,2n-6,\ldots$.
The terms of (\ref{popu2})
with $\sum_j n_j=2n$ are clearly related to reducible graphs,
the other ones are due to the subtlety mentioned above and do not
contribute to irreducible graphs.
Since $\ll 1\gg_c=F(0,t_0,t_1,\ldots)\equiv F$, we conclude
\begin{equation}
<\prod_{i=0}^\infty\sigma_i^{k_i}>^{irr}_n=
{1\over n!}\left({1\over 2}{\partial^2\over \partial t_0^2}\right)^n
\left.\prod_{i=0}^\infty{\partial^{k_i}\over \partial t_i^{k_i}}
F(0,t)\right|_{t=0}=<{1\over n!}\left({\sigma_0\sigma_0\over 2}
\right)^2\prod_{i=0}^\infty\sigma_i^{k_i}>_0,
\end{equation}
as desired. Let examine the remaining terms of
(\ref{popu2}). Their meaning
will be illustrated with
the explicit examples $n=1,2,3$.
For $n=1$ we have
\begin{equation}
<\ldots >_1={1\over 2}F_{xx}
+{1\over 2}F_xF_x,
\label{n11}
\end{equation}
where the subscripts denote pairwise identified $t_0$-derivatives.

It is evident that the second term collects contributions from
reducible graphs, while there is no ``puzzling'' term.
Each derivative of $F$ represents a component.
For $n=2$ we have
\begin{equation}
<\ldots >_2={1\over 2!}{1\over 2^2}
F_{xxyy}+{1\over 2}F_{xxy}F_y
+{1\over 4}F_{xy}F_{xy}+{1\over 2}F_xF_{xy}F_y
-{1\over 2}{\partial F\over \partial t_0}.
\label{n2}
\end{equation}
The ``puzzling'' term is the last one.
Let us define
\begin{equation}
{\partial^3 F^\prime \over \partial t_0^3}=
{\partial^3 F\over \partial t_0^3}-1.
\label{repla}
\end{equation}
The puzzling term is reabsorbed by replacing the third derivative of
$F$
with the third derivative of $F^\prime$.
That this works in general is promptly checked for $n=3$:
\begin{eqnarray}
<\ldots>_3&=&{1\over 3!}{1\over 2^3}F_{xxyyzz}
+{1\over 2^3}F_{xxyyz}F_z+{1\over 2^2}F_{xxyz}F_{yz}
+{1\over 2^3}F^\prime_{xxy}F^\prime_{yzz}
+{1\over 3!}{1\over 2}F^\prime_{xyz}F^\prime_{xyz}
\nonumber\\&&
+{1\over 2^2}F_{xxyz}F_yF_z
+{1\over 2}F^\prime_{xxy}F_{yz}F_z
+{1\over 2}F_{xy}F^\prime_{xyz}F_z+
{1\over 3!}F_{xy}F_{yz}F_{zx}\nonumber\\&&
+{1\over 3!}F^\prime_{xyz}F_xF_yF_z
+{1\over 2}F_xF_{xy}F_{yz}F_z.
\label{n3}
\end{eqnarray}
One can check that the coefficients of each one of the above terms
is the correct symmetry factor. It is amazing to notice that the
operator
$(\tr[X])^2$ describes all kinds of nodes at the same time.

The reason while one has to subtract $1=
\left.{\partial^3 F\over \partial t_0^3}\right|_{t=0}$ from
${\partial^3 F\over \partial t_0^3}$ is
that the graphs collected by the matrix model
(\ref{compmodel}) are such that in any component there should be
at least one
marked point (beyond those due to the nodes).
The $t_0$-derivatives appearing in (\ref{n11}),
(\ref{n2}) and (\ref{n3})
correspond to nodes. Without the replacement (\ref{repla})
there would be terms with components having three virtual
marked points and no true marked point. This would correspond
to a genus $0$ ribbon graph with
three $\tr[X]$-insertions and no face, i.e.\
a quadratic differential $\varphi$ on the sphere
with three simple poles only,
which is impossible.

Finally, the reason why any component should have at least one
true marked point is easily found.
In the proof of the theorem of section \ref{nodes}, I associated
virtual faces to the marked points describing nodes
and let the virtual perimeters tend to zero.
The privileged configuration is the one in which every virtual face
is surrounded by a single link. One can easily convince oneself
that it is impossible to surround {\sl all}
the marked points in this way: one needs at least one marked point
surrounded by a polygon with more links. This is
the required ``true'' face.

Formula (\ref{equa}) is only one particular case of the theorem
proved in
\cite{difra}. The general structure of
formula (\ref{yto}) is very similar to formula (\ref{popu}),
in the sense that the polynomial $Q$ of (\ref{yto})
is the sum of a first term that is
apparently reminiscent of $P(X)$ plus lots of complicated extra terms
with ``lower weights''\footnotemark
\footnotetext{For the precise statement, see \cite{cornalba}.
The ``lower weights'' differ from the ``highest weight'' by
multiples of three units, as in (\ref{popu}).}.
Perhaps, geometrical interpretations of those extra terms can also
be found.

In ref.\ \cite{cornalba} the conjectured relation between
the sets ${\cal M}_{m*,s}$ and polynomials in the Mumford-Morita
classes
is investigated.
The results of the present paper
give the interpretation of $m_0$ in terms of the trivial
``Mumford-Morita class'' $1$.

\section{Conclusions}
\label{concl}
\setcounter{equation}{0}

Intersection theory on surfaces with nodes can be seen as a
kind of ``constrained topological gravity'' \cite{preprint,letter},
namely a theory of topological gravity in which the moduli space is
some proper submanifold (a cocycle) of the moduli space ${\cal M}_g$
of Riemann surfaces of genus $g$.
One can thus get a {\sl finer} look at the moduli space.
In the present case the constraint is represented by nodes.
The pinching of a non-separating node is usually denoted by
$\Delta_0$ and is indeed a cocycle. The pinching of $n$
non-separating nodes
is $\Delta_0\cap\Delta_0\cap\ldots $($n$ times).
In \cite{preprint,letter} the proper submanifold
was the Poincar\`e dual of the top Chern class $c_g({\cal E}_{hol})
$ of the Hodge bundle ${\cal E}_{hol}\rightarrow {\cal M}_g$.
This constraint is indeed easily realizable in a quantum field
theoretical model \cite{letter} and is directly suggested by
N=2 supersymmetry \cite{preprint}.
The idea is that
of looking for a finer investigation of the moduli
space of a topological field theory by letting the gauge-fixing BRST
algebra contain fields (the Lagrange multipliers in particular)
with nontrivial global degrees of freedom. These moduli produce in a
natural
way ``BRST constraints'', instead of enlarging the moduli space.
A moduli space constraint could permit to satisfy selection rules
that cannot
be satisfied
within the usual class of topological observables.
A precise identification of the surfaces lying in the Poincar\'e dual
${\cal V}_{g,s}$ of $c_g({\cal E}_{hol})$ is still not available
in the mathematical literature and it is worth
considering the possibilitites that are suggested by physics, in
order
to reach a deeper knowledge on
the moduli space of Riemann surfaces.
It would be very interesting to characterize ${\cal V}_{g,s}$  in
terms of a constraint on the set of ribbon graphs.
For this purpose, it could be important to possess the ``graphical
counterparts'' of certain ``geometrical notions'', extending
the discussion of section \ref{nodes} that dealt with the graphical
counterpart of the plumbing fixture.
In \cite{preprint} it was noticed that,
if the Poincar\'e dual ${\cal V}_{g,s}$
of $c_g({\cal E}_{hol})$ has a representative in the boundary of
${\cal
M}_{g}$, the simplest possibility is that it is the set of
completely degenerate Riemann
surfaces, those with all the A-cycles pinched. This set
is equivalent to the set of spheres with $g$ couples of
pairwise identified points.
It this conjecture
is correct, then formula (\ref{for}) gives,
with $n$ replaced by $g$,
all the correlation functions of the model of ref.s
\cite{preprint,letter}.
Formula (\ref{argue}) could then be expressed as
\begin{equation}
c_g({\cal E}_{hol})\sim\Delta_0^g\sim{1\over g!}\left({\sigma_0
\sigma_0\over 2}\right)^g.
\label{po}
\end{equation}
Now, $c_g({\cal E}_{hol})$ is a {\sl nonlocal} class,
differently from the Mumford-Morita classes. Then, formula
(\ref{po}) expresses the fact that the {\sl nonlocality}
is fully encoded in the factor ${1\over g!}$. Indeed,
such factor requires the knowledge of the
total number of couples of identified points.

As a matter of fact, the
conjecture is true at least for $g=1$, where ${\cal V}_{g,s}$ is
a point and can be chosen to be the singular torus. Formula
(\ref{for}) for $g=1$ matches with the correlation functions in genus
one that were given at the end of \cite{letter}.
However, the general proof does not seem straightforward.
The situation is now that we have a field theoretical model for
$c_g({
\cal E}_{hol})$ and a matrix model for nodes. One would also
like to formulate a field theoretical model for nodes and a matrix
model for $c_g({\cal E}_{hol})$. Then, comparison of the two is
expected to be a source of insight in the study of the
moduli space of Riemann
surfaces, permitting to prove, improve or disprove the conjecture.

The quadratic and ``non propagating'' operator $(\tr[X])^2$ may be
relevant for other applications in matrix models. In \cite{das}
the composite operator $(\tr[X^2])^2$ is discussed in detail, viewed
as an effect of higher order curvature terms. Three different phases
are present in the model of \cite{das}: smooth surfaces, branched
polymers and an intermediate phase.
The simpler operator $(\tr[X])^2$ could play some interesting role in
similar contexts.

\vspace{12pt}
\begin{center}
{\bf Acknowledgements}
\end{center}
\vspace{6pt}

I would like to thank C.\ Reina, P.\ Fr\`e and M.\ Matone for useful
discussions.

\end{document}